%% file: ms.tex
\documentclass[sigconf]{acmart}
\usepackage{booktabs} 
\usepackage{multirow}
\usepackage{color}
\usepackage{subfigure} 
\usepackage{array}
\usepackage{inputenc}
\usepackage{tabularx}
\usepackage{longtable}
\usepackage{booktabs}
\usepackage{algorithm}
\usepackage{algorithmic}
\usepackage{graphicx}
\usepackage{enumitem}
\usepackage{amsfonts}
\usepackage{hyperref}
\hypersetup{colorlinks=false,linkcolor=blue,urlcolor=blue,citecolor=red}

\newcommand{\Lapl}{\mathbf{\mathop{\mathcal{L}}}}
\newcommand{\Trans}[1]{{#1}^{\top}}

\newcommand{\Mat}[1]{\mathbf{#1}}

\newcommand{\Space}[1]{\mathbb{#1}}
\newcommand{\Set}[1]{\mathcal{#1}}

\newcommand{\ie}{\emph{i.e., }}
\newcommand{\eg}{\emph{e.g., }}

\newcommand{\wrt}{\emph{w.r.t. }}
\newcommand{\cf}{\emph{cf. }}

\theoremstyle{definition}

\AtBeginDocument{%
  \providecommand\BibTeX{{%
    \normalfont B\kern-0.5em{\scshape i\kern-0.25em b}\kern-0.8em\TeX}}}

\copyrightyear{2020}
\acmYear{2020}
\setcopyright{iw3c2w3}
\acmConference[WWW '20]{Proceedings of The Web Conference 2020}{April 20--24, 2020}{Taipei, Taiwan}
\acmBooktitle{Proceedings of The Web Conference 2020 (WWW '20), April 20--24, 2020, Taipei, Taiwan}
\acmPrice{}
\acmDOI{10.1145/3366423.3380098}
\acmISBN{978-1-4503-7023-3/20/04}

\copyrightyear{2020}
\acmYear{2020}
\setcopyright{iw3c2w3}
\acmConference[WWW '20]{Proceedings of The Web Conference 2020}{April 20--24, 2020}{Taipei, Taiwan}
\acmBooktitle{Proceedings of The Web Conference 2020 (WWW '20), April 20--24, 2020, Taipei, Taiwan}
\acmPrice{}
\acmDOI{10.1145/3366423.3380098}
\acmISBN{978-1-4503-7023-3/20/04}

\begin{document}

\settopmatter{printacmref=true}

\title{Reinforced Negative Sampling over Knowledge Graph for Recommendation}

\author{Xiang Wang}
\affiliation{%
	\institution{National University of Singapore}
}
\email{xiangwang@u.nus.edu}

\author{Yaokun Xu}
\affiliation{%
	\institution{Southeast University}
}
\email{xuyaokun98@gmail.com}

\author{Xiangnan He}
\affiliation{%
	\institution{University of Science and Technology of China}
}
\email{xiangnanhe@gmail.com}

\author{Yixin Cao}
\affiliation{%
	\institution{National University of Singapore}
}
\email{caoyixin2011@gmail.com}

\author{Meng Wang}
\affiliation{%
	\institution{HeFei University of Technology}
}
\email{eric.mengwang@gmail.com}

\author{Tat-Seng Chua}
\affiliation{%
	\institution{National University of Singapore}
}
\email{dcscts@nus.edu.sg}

\begin{abstract}
Properly handling missing data is a fundamental challenge in recommendation. Most present works perform negative sampling from unobserved data to supply the training of recommender models with negative signals. Nevertheless, existing negative sampling strategies, either static or adaptive ones, are insufficient to yield high-quality negative samples --- both informative to model training and reflective of user real needs.

In this work, we hypothesize that item knowledge graph (KG), which provides rich relations among items and KG entities, could be useful to infer informative and factual negative samples.
Towards this end, we develop a new negative sampling model, \emph{Knowledge Graph Policy Network} (KGPolicy), which works as a reinforcement learning agent to explore high-quality negatives.
Specifically, by conducting our designed exploration operations, it navigates from the target positive interaction, adaptively receives knowledge-aware negative signals, and ultimately yields a potential negative item to train the recommender.
We tested on a matrix factorization (MF) model equipped with KGPolicy, and it achieves significant improvements over both state-of-the-art sampling methods like DNS~\cite{DNS} and IRGAN~\cite{IRGAN}, and KG-enhanced recommender models like KGAT~\cite{KGAT}.
Further analyses from different angles provide insights of knowledge-aware sampling.
We release the codes and datasets at \url{https://github.com/xiangwang1223/kgpolicy}.

\end{abstract}

\ccsdesc[500]{Information systems~Recommender systems}
\keywords{Recommendation, Knowledge Graph, Negative Sampling}

\maketitle

\input{1_introduction.tex}

\input{2_preliminary.tex}
\input{3_method.tex}
\input{4_experiment_correct.tex}
\input{5_related.tex}
\input{6_conclusion.tex}

\bibliographystyle{ACM-Reference-Format}
\balance
\bibliography{ms}
\balance


\end{document}

%% file: 1_introduction.tex
\section{Introduction}
Recommender systems have been widely adopted in real-world applications to improve user satisfaction and engagement.
To train a recommender model from historical user-item interactions, both positive and negative user feedback are required to ensure that the model generates reasonable personalized ranking~\cite{BPRMF,eALS,NSCR}. 
Nevertheless, most interactions are in the form of implicit feedback, \eg clicks and purchases, which provide only the signals on positive feedback.
This brings in the fundamental challenge to recommender model learning --- how to distill negative signals from the positive-only data --- which is also known as the one-class problem~\cite{OCCF}. 

\begin{figure}[t]
    \centering
	\includegraphics[width=0.45\textwidth]{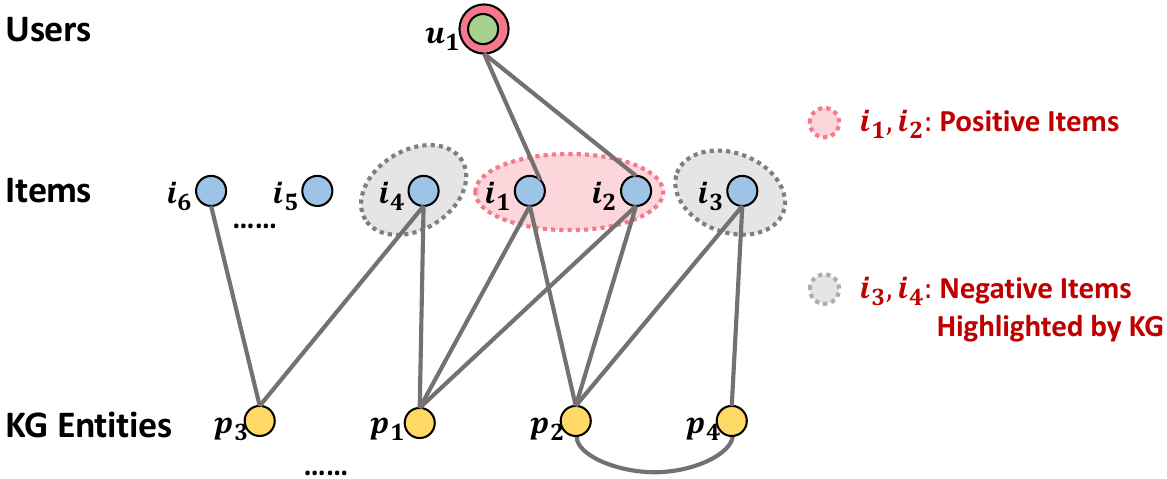}
	\vspace{-10pt}
	\caption{A toy example of distilling negative signals from KG. Having overlapping KG entities with $i_{1}$ and $i_{2}$, $i_3$ and $i_4$ are more likely to be of less interest to $u_1$.
	}
	\label{fig:intro}
	\vspace{-10pt}
\end{figure}

As negative signals are latent in unobserved data, a prevalent solution is to perform negative sampling~\cite{BPRMF}, which is more efficient and versatile than treating all unobserved interactions as negative~\cite{OCCF,eALS}. Existing strategies on negative sampling can be categorized into three types --- static sampler~\cite{BPRMF,RNS,PNS}, adaptive sampler for hard negatives~\cite{AdaptiveSampler,LambdaFM,DNS,IRGAN}, and enhanced sampler with extra behaviors~\cite{RNS,DBLP:conf/recsys/LoniPLH16}. However, each type of method suffers from some inherent limitations:
\begin{itemize}[leftmargin=*]
    \item Static sampler applies fixed distribution to sample from missing data, \eg uniform~\cite{BPRMF,NCF} and popularity-biased distribution~\cite{PNS,DBLP:conf/kdd/ChenSSH17}. The main limitation is that it is independent of model status, making it easier to yield low-quality negative samples. For example, if a user-item instance is already scored lowly by the current model, sampling it as a negative example will have minor change on model parameters (as gradients are close to zero~\cite{AdaptiveSampler}).   
    \item Adaptive sampler prefers hard negatives, since training on them can bring large change on current model parameters~\cite{DNS,AdaptiveSampler,IRGAN}. For example, DNS~\cite{DNS} picks the one scored highest by the current model among some random unobserved samples, and IRGAN~\cite{IRGAN} optimizes this process under a generative adversarial framework. While being effective from the perspective of numerical optimization, these hard negatives are likely to be true positives (in future testing data), which degrades model effectiveness. 
    \item Some recent works incorporate extra behavior data, \eg viewed but non-clicked~\cite{RNS,DBLP:conf/www/DingF0YLJ18} and clicked but non-purchased~\cite{DBLP:conf/recsys/LoniPLH16}, to enhance the negative sampler. Although such data provides certain signal on true negatives, it is of limited scale as compared with the vast amount of missing data. As such, only using them as negative feedback is rather insufficient (even performs worse than the uniform sampler~\cite{DBLP:conf/www/DingF0YLJ18}). For this type of method, it still needs strategies to effectively distill negative signal from the massive missing data. 
\end{itemize}

Given the fundamental role of negative sampling and the limitations of existing methods, we focus on negative sampling in this work, aiming to improve its quality by introducing other types of data. 
We claim that high-quality negative samples should satisfy two requirements: 1) \textit{informative}, meaning that the current model rates them relatively high, such that updating them as negative feedback will change model parameters significantly, and 2) \textit{factual}, meaning that they are true negatives, \ie the user has known them before (exposed by the system or through other ways) but did not choose them. As the requirement of informative can be achieved by adaptive sampler, the key challenge lies in discovering factual negatives from missing data, which lacks ground-truth by nature and has not been well addressed by previous work.

In this work, we hypothesize that knowledge graph (KG), which introduces extra relations among items and real-world entities (from item attributes or external knowledge), could be useful to infer factual negative samples from unobserved data.
Although incorporating KG into recommendation has recently been extensively researched~\cite{KGAT,CKE,RippleNet}, these studies only leverage KG to build the predictive model, and none of previous works has considered using it to enhance the negative sampler.
In particular, they assume that items, which have overlapping KG entities with historical items, would be unexposed but of interest to the target user.
However, in real-world scenarios, a user is often aware of these items through some ways (\eg searching, mouth marketing, or advertising systems); hence, she does not adopt them, suggesting that she might be truly not interested in these items.
An example is shown in Figure~\ref{fig:intro}, where user $u_1$ watches movies $i_1$ and $i_2$, both of which are directed by the same person $p_1$ and of the same genre $p_2$.
We may infer that the combination of director $p_1$ and genre $p_2$ is an important factor of $u_1$'s interest.
Thus $u_1$ is highly likely to have known other movies (\eg $i_4$) directed by $p_1$ but with different genres, but be less interested on them.
As such, from the perspective of negative sampling, $i_{4}$ could offer high-quality negative signals.

Despite great potential, it is highly challenging to exploit KG to guide negative sampling. First, when exploring KG towards possible negative items, the scale of exploring paths (\eg $\{i_1,i_2\}\rightarrow p_{1}\rightarrow i_{4}$ in Figure~\ref{fig:intro})~increases dramatically, since many edges are continually added at each forward step. 
Thus, there is a strong need for an intelligent sampler to effectively traverse KG.
Second, due to the lack of ground-truth, it requires the sampler to distinguish negative signals carried by the exploring paths --- more specifically, differentiate the confidence of KG entity (\eg $p_1$) being exposed to the target user and estimate the probability of possible item (\eg $i_4$) being negative.




Towards this end, we propose a new negative sampling model, KGPolicy (short for \textit{Knowledge Graph Policy Network}), which employs a reinforcement learning (RL) agent to explore KG to discover high-quality negative examples.
At the core is the designed \textbf{exploration operation}, which navigates from the positive item, picks two sequential neighbors (\eg one KG entity and one item) to visit.
Such a two-hop path captures the knowledge-aware negative signals.
We devise a neighbor attention module to achieve this goal, which specifies varying importance of one- and two-hop neighbors conditioned on the positive user-item pair, so as to adaptively capture personal tastes on KG entities and yield potential items.
By conducting such exploration recursively, KGPolicy learns to select potential negative items for a target positive interaction.
Moreover, the path history works as a support evidence revealing why the selected item is being treated as a negative instance. 
To demonstrate our method, we employ a simple linear model, matrix factorization (MF), as the recommender and co-train it with our KGPolicy.
Empirically, MF equipped with KGPolicy achieves significant improvements over both state-of-the-art sampling methods like DNS~\cite{DNS} and IRGAN~\cite{IRGAN}, and KG-enhanced recommender models like KGAT~\cite{KGAT}, on three benchmark datasets.
Further analyses provide insights on how knowledge-aware negative samples facilitate the learning of recommender \wrt two requirements --- informative (evidence in the training process \wrt gradient magnitude, and the performance \wrt sparsity level) and reflective of personal tastes (evidences in case study).
It is worth highlighting that KGPolicy is recommender-agnostic and can works as a plug-and-play sampler for arbitrary recommenders.

In a nutshell, this work makes the following main contributions:
\begin{itemize}[leftmargin=*]
    \item To the best of our knowledge, we are the first to incorporate knowledge graph into negative sampling, with the aim of selecting high-quality negative samples to pair with a positive user-item interaction.
    \item We develop a reinforcement learning agent for negative sampling, KGPolicy, which effectively learns to navigate towards high-quality negative items with multi-hop exploring paths.
    \item We conduct extensive experiments on three benchmark datasets, demonstrating the advantages of KGPolicy on the effectiveness of sampling and the usage of knowledge entries.
\end{itemize}

%% file: 2_preliminary.tex
\section{Task Formulation}\label{sec:task-formulation}
We first present interaction data and knowledge graph, formulate our task, and emphasize negative signals within multi-hop paths.

\begin{figure*}[t]
    \centering
	\includegraphics[width=0.82\textwidth]{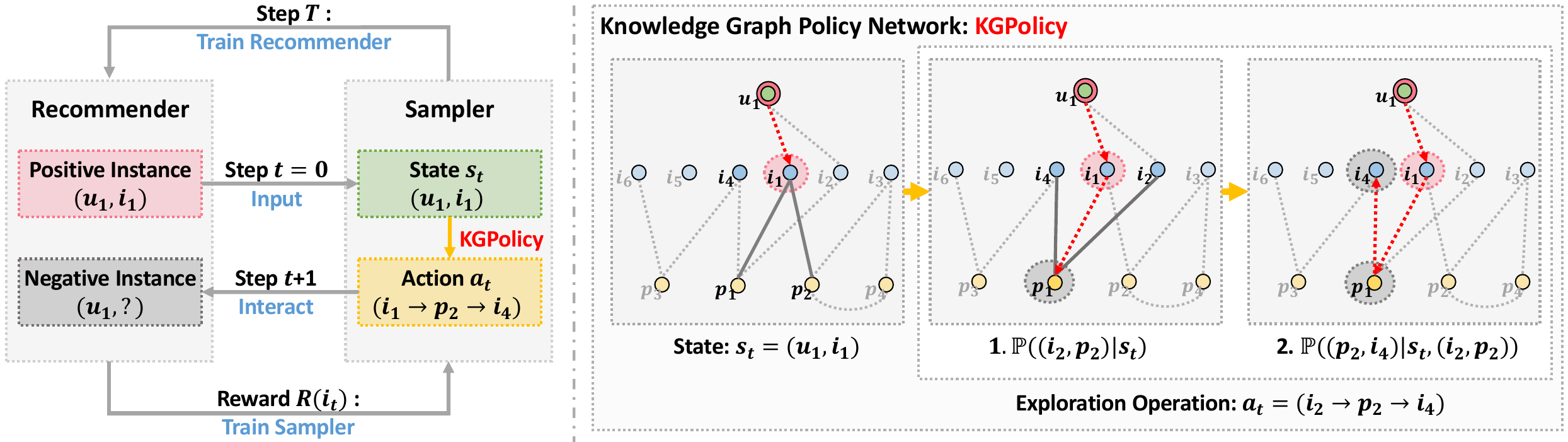}
	\vspace{-10pt}
	\caption{Illustration of the proposed knowledge-aware negative sampling framework. The left subfigure shows the model framework, and the right subfigure presents the proposed KGPolicy network. Best view in color.}
	\label{fig:framework}
	\vspace{-10pt}
\end{figure*}


\vspace{5pt}
\noindent\textbf{Interaction Data.}
Let $\Set{O}^{+}=\{(u,i)|u\in\Set{U},i\in\Set{I}\}$ be the implicit feedback, where each $(u,i)$ pair indicates a historical interaction between user $u$ and positive item $i$, and $\Set{U}$ and $\Set{I}$ denote the sets of users and items, respectively.


\vspace{5pt}
\noindent\textbf{Knowledge Graph.}
Inspired by recent works~\cite{KTUP,KGAT}, we organize item attributes or external knowledge as well as interaction data in the form of knowledge graph (KG).
As prior efforts~\cite{KB4Rec} show, items in user-item interaction data can be aligned with the corresponding entities in KG.
With such alignments, we can establish a KG, formalized as $\Set{G}=\{(e,e')|e,e'\in\Set{E}\}$, where $\Set{E}$ is the entity set unifying the user and item sets $\Set{U}$ and $\Set{I}$ with KG entity set $\Set{P}$, \ie $\Set{E}=\Set{U}\cup\Set{I}\cup\Set{P}$. For example, (\emph{item: Harry Potter}, \emph{author: J. K. Rowling}) describes the fact that \emph{J. K. Rowling} is the author of the book \emph{Harry Potter}. Here we simplify the entity-relation-entity KG triples as the edges between conceptualized entities, merging the semantic relations into the entity instances and leaving the explicit modeling of them in future work.

\vspace{5pt}
\noindent\textbf{Task Description.}
Having established user behaviors and item knowledge, we aim to exploit such rich information to guide the sampler learning.
Our goal is the knowledge-aware negative sampling, formulated as:
\begin{align}
    j\sim f_{S}(u,i,\Set{G}),
\end{align}
where $f_{S}(\cdot)$ is the sampler parameterized with $\Theta_{S}$.
It generates the empirical distribution over unobserved items to yield knowledge-aware negative item $j$, which is expected to be informative and reflective of personal tastes.
Furthermore, the exclusive KG entities of positive $i$, $\{p|(i,p)\in\Set{G}, (j,p)\notin\Set{G}\}$, helps explain why the target user $u$ is less interested on negative $j$.
For example, $\{p_{2}\}$ might be the reason on $u_1$'s behavior difference between $i_{1}$ and $i_{4}$.


\vspace{5pt}
\noindent\textbf{Negative Signals in Multi-hop Paths.}
Towards that, we aim to explore the structural information of KG, especially high-order connectivity among nodes, to discover suitable negative items.
For a positive $(u,i)$ interaction, we can traverse paths rooted at node $i$, terminate at a unobserved item $j$, and view multi-hop connections as the relationships between $i$ and $j$.
However, it is obvious that various paths have different confidences in the discovered items being negative, and not all paths are useful for distilling negative signals.
We hence heuristically define the atomic path type as:
\begin{itemize}
    \item $i \rightarrow e' \rightarrow j$ with $e'\in\Set{E}$,
\end{itemize}
which is 1) \textbf{informative}, since two items $i$ and $j$ share the same KG entity $e'$, they might have similar representations and their pairwise comparison might provide large gradients on the recommender parameters; and 2) \textbf{reflective of user real tastes}, since if $e'$ is an important factor of $u$'s interest, $j$ might have been exposed to $u$ through other means (\eg searching, mouth marketing, or advertising systems). However, $u$ consumed $i$ rather than $j$, suggesting that she might be truly less interested in $j$ compared to $i$.
As a result, $(u,j)$ is expected as a better negative sample to train the recommender.
Moreover, if $j$ is estimated with lower confidence being negative, we can continue the exploration by extending such atomic paths.
For example, item $j'$ can be discovered from $i \rightarrow e \rightarrow j' \rightarrow e' \rightarrow j$ with higher confidence of being negative.

%% file: 3_method.tex
\section{Methodology}
We now present knowledge-aware negative sampling. Figure~\ref{fig:framework} shows the framework, which consists of one recommender and the proposed sampler.
We then elaborate our sampler, KGPolicy, with the target of learning to navigate towards informative negatives on KG.
Specifically, there are three main components conducting the exploration operations: 1) graph learning module, which prepares high-quality representations of nodes in advance; 2) neighbor attention module, which utilizes two attention models to conduct path finding and determines which suitable node to visit next; and 3) neighbor pruning module, which reduce the search space to solve the computational overload in the foregoing module.
Recursively performing such explorations, KGPolicy ultimately is able to yield a potential item to pair the positive target.
Finally, KGPolicy and the recommender are co-trained for recommendation.

\subsection{Recommender}
To demonstrate the effectiveness of our knowledge-aware sampler, we employ a linear and simple model, matrix factorization (MF)~\cite{BPRMF}, as the recommender.
To be more specific, MF parameterizes ID information of users and items as embeddings, and uses inner product of user and item embeddings as the predictive function to estimate how likely user $u$ would consume item $i$.
The holistic goal is formulated as:
\begin{align}\label{equ:mf-recommender}
    \hat{y}_{ui}=f_{R}(u,i)=\Trans{\Mat{r}}_{u}\Mat{r}_{i},
\end{align}
where $\hat{y}_{ui}$ is the prediction score for an $(u,i)$ interaction; $f_{R}(\cdot)$ is abstracted as the interaction function with recommender parameters $\Theta_{R}$; $\Mat{r}_{u}\in\Space{R}^{d}$ and $\Mat{r}_{i}\in\Space{R}^{d}$ are ID embeddings of user $u$ and item $i$, respectively; $d$ is the embedding size.

Following prior studies~\cite{NCF,BPRMF,NGCF}, we use the pairwise BPR loss~\cite{BPRMF} as the objective function to optimize and learn the parameters $\Theta_{R}$. 
Specifically, it assumes that, for a target user, her historical items reflecting more personal interest should be assigned higher prediction scores, than that of unobserved items, as:
\begin{align}\label{equ:def-recommender-loss}
    \min_{\Theta_{R}}\sum_{(u,i)\in\Set{O}^{+}}\Space{E}_{j\sim f_{S}(u,i,\Set{G})}-\ln{\sigma\Big(f_{R}(u,i)-f_{R}(u,j)\Big)},
\end{align}
where $\sigma(\cdot)$ is the sigmoid function.
As such, the recommender as a critic judges whether the items $i$ and $j$ are truly consumed by user $u$.
Moreover, following previous work~\cite{AdaptiveSampler}, the informativeness of a negative sample can be measured as the gradient magnitude, as:
\begin{align}\label{equ:gradient-magnitude}
    \Delta_{u,i,j}=1-\sigma(f_{R}(u,i)-f_{R}(u,j)),
\end{align}
which reflects the contribution a pairwise preference $(u,i,j)$ has for improving $\Theta_{R}$.
Low-quality negatives, which are assigned a smaller score than $i$, make gradient magnitude close to $0$, hence contribute little to optimization. 
Hence, an informative negative is expected to have close prediction scores with the positive target.

\subsection{Knowledge-aware Sampler}
While adaptive samplers~\cite{AdaptiveSampler,DNS,LambdaFM} achieve great success towards informative negatives, the discovery of factual negative signals are not fully explored.
Towards this end, we take KG as the priors (or environment) of the sampler.
This allows us to exploit rich relations, especially high-order connectivity, among items and KG entities for exploring more suitable negatives.
As shown in the right side of Figure~\ref{fig:framework}, the basic idea is to, conditioned on the target user, start from the positive item, learn to navigate over the KG structure, then yield the possible negatives along the exploring paths.
Such paths are composed of the atomic paths defined in Section~\ref{sec:task-formulation}.

Enumerating possible paths towards all unobserved items is infeasible in large-scale KGs, since it requires labor-intensive feature engineering, being memory-consuming to store these paths and time-consuming to distill useful signals.
Thus, we design an intelligent sampler as a reinforcement learning (RL) agent to conduct automatic exploration over KG.

\subsubsection{\textbf{Sampling as Reinforcement Learning.}}
We cast sampling as a Markov Decision Process (MDP) $M=\{\Set{S},\Set{A},\Set{P},\Set{R}\}$, where $\Set{A}=\{a\}$ is the set of actions derived from exploration operations, $\Set{S}=\{s\}$ is the set of states abstracting paths during exploration, $\Set{P}$ is the transition dynamics of states, and $\Set{R}$ is a reward function.
We introduce these key elements of KG environment for RL as follows: 

\vspace{5pt}
\noindent\textbf{Exploration Operation.}
To obtain the atomic paths defined in Section~\ref{sec:task-formulation}, we define a new exploration operation involving two successive edges.
Formally, at step $t$, $a_{t}=(e_{t}\rightarrow e'_{t}\rightarrow e_{t+1})$ denotes a two-hop path rooted at item $e_{t}$ towards a proposal $e_{t+1}$, where $(e_t,e'_t)$, $(e'_t,e_{t+1})\in\Set{G}$, and $e_{t}$, $e_{t+1}\in\Set{I}$ are connected via the internal node $e'_{t}$.
Such operation considers the confidence of the KG entity $e'_{t}$ being exposed to the target user $u$, and the confidence of item $e_{t+1}$ being negative.
With the budge of $T$ exploration operations, the sampler $f_{S}(\cdot)$ generates a $2T$-hop paths $(a_{1},\cdots,a_{T})$, which sequentially yields $T$ proposals as a policy:
\begin{align}\label{equ:policy}
    \pi=(e_{1},e_{2},\cdots,e_{T}),
\end{align}
where the positive item $i$ is the first node in $a_{0}$; $e_{t}$ is the last item node of the $t$-th step.
At the terminal step $T$, $e_{T}$ is used as the final negative item to optimize the recommender.
Varying the number of exploration operations $T$ allows us to flexibly adjust the search space, so as to guarantee the diversity of negatives.
We set $T$ as $2$ by default and evaluate the impact of $T$ in Section~\ref{sec:impact-k}.

\vspace{5pt}
\noindent\textbf{State.}
At step $t$, the state $s_{t}$ conditioned on the target user $u$ is defined as a triple $(u,e_{t})$, where $e_{t}$ is the node the sampler visits currently.
As such, we can formalize the exploration process prior to step $t$ as $\{s_{0},a_{1},s_{1},\cdots,a_{t},s_{t}\}$, where the initial state $s_{0}$ is $\{u,i\}$.

\vspace{5pt}
\noindent\textbf{Action.}
The action space $\Set{A}_{t}$ of state $s_{t}$ is composed of all possible exploration operations starting from node $e_{t}$, excluding the historical trajectory.
As the state is changing during the exploration and its neighbors are different, the action space is dynamic.

\vspace{5pt}
\noindent\textbf{State Transition Dynamics.}
Given an exploration operation $a_{t}$ at state $s_{t}$, the transition to the next state $s_{t+1}$ is determined as:
\begin{align}
    \Space{P}\Big(s_{t+1}=(u,e_{t+1})|s_{t}=(u,e_{t}), a_{t}=(e_{t}\rightarrow e'_{t}\rightarrow e_{t+1})\Big)=1.
\end{align}

\vspace{5pt}
\noindent\textbf{Reward.}
The reward $\Set{R}_{e_{t}}$ measures the quality of the proposal item $e_{t}$ \wrt the pairwise preference $(u,i,j)$ at step $t$.
However, without ground truth of negative signals to confirm whether $u$ is truly less interested on $e_{t}$, we rely on the feedback from the recommender to define the soft reward function.
Here we consider two factors:
\begin{itemize}[leftmargin=*]
    \item\textbf{Prediction Reward:} the prediction of negative $e_{t}$ is the typical reward in adversarial sampling methods~\cite{IRGAN,AdvIR}, accounting for the matching score between $e_{t}$ and $u$. The sampler is encouraged by the recommender to yield items with higher predictions. From the perspective of informativeness, an item ranked close to the positive is able to offer larger gradient magnitude.
    \item\textbf{Similarity Reward:} intuitively, if $e_{t}$ is similar to positive item $i$, $e_{t}$ is more likely to be exposed to user $u$. The recommender enforces the sampler to care exposed but less interested items.
\end{itemize}
Considering these two factors, we design a reward function as:
\begin{align}\label{equ:reward}
    \Set{R}(e_{t})=f_{R}(u,e_{t})+g_{R}(i,e_{t}),
\end{align}
where $f_{R}(u,e_{t})=\Trans{\Mat{r}_{u}}\Mat{r}_{e_{t}}$ and $g_{R}(i,e_{t})=\Trans{\Mat{r}_{i}}\Mat{r_{e_{t}}}$ separately denote prediction and similarity rewards;
$\Mat{r}_{u}$, $\Mat{r}_{i}$, and $\Mat{r}_{e_{t}}$ are the recommender's ID embeddings of $u$, $i$, and $e_{t}$, respectively.
Here we set equal importance for the two reward components, leaving their linear combination controlled by the hyper-parameter future.
We verify the rationality of the two reward functions in Section~\ref{sec:impact-reward}.

\vspace{5pt}
\noindent\textbf{Objective Function.}
Towards learning a stochastic policy $\pi$ to optimize the sampler parameters $\Theta_{S}$, we maximize the expected cumulative discounted reward as follows:
\begin{align}
    \max_{\Theta_{S}}\sum_{(u,i)\in\Set{O}^{+}}\Space{E}_{\pi}[\sum_{t=1}^{T}\lambda^{t-1}\Set{R}(e_{t})],
\end{align}
where $\gamma$ is the decay factor;
the expectation of $\pi$ is to make the likelihood of a proposal pair as close to that of the possible interaction as possible.
In what follows, we elaborate our policy network towards obtaining the probability of $e_{t}$ being negative in $\pi$, \ie $\Space{P}(a_{t}|s_{t})$, at $t$-th step.

\subsection{\textbf{Knowledge Graph Policy Network.}}
In this section, we introduce a network to generate policy $\pi$, as well as the confidence for each action. 
First, we describe a graph learning module, which generates representation for each node, and then build a neighbor attention module upon the representations to pick a suitable neighbor as a proposal to visit, which is coupled with a neighbor pruning module to reduce the exploration space.


\subsubsection{\textbf{Graph Learning Module.}}\label{sec:graph-learning-module}
Inspired by recent graph neural networks (GNNs)~\cite{GCN,GAT,GraphSAGE,NGCF} which are powerful to generate representations for graph data, we employ GraphSage~\cite{GraphSAGE} on $\Set{G}$ and the user-item bipartite graph $\Set{O}^{+}$, to embed user, item, and KG entity nodes.
In particular, at the $l$-th graph convolutional layers, a node $e$ receives the information being propagated from its neighbors to update its representation, as:
\begin{align}\label{equ:gnn-rep}
    \Mat{h}^{(l)}_{e}=\rho\Big(\Mat{W}^{(l)}(\Mat{h}^{(l-1)}_{e}||\Mat{h}^{(l-1)}_{\Set{N}_{e}})\Big),
\end{align}
where $\Mat{h}^{(l)}_{e}\in\Space{R}^{d_{l}}$ is the representation after $l$ steps of embedding propagation, $d_{l}$ denotes the embedding size, $\Mat{W}^{(l)}\in\Space{R}^{d_{l}\times 2d_{l-1}}$ is the weight matrix to distill useful information; $||$ is the concatenation operation and $\rho$ is a nonlinear activation function set as LeakyReLU~\cite{GraphSAGE,GAT} here; $\Mat{h}^{(0)}_{e}$ represents ID embedding, while $\Mat{h}^{(l-1)}_{\Set{N}_{e}}$ is the information being propagated from $e$'s neighbors as:
\begin{align}
    \Mat{h}^{(l-1)}_{\Set{N}_{e}}=\sum_{e'\in\Set{N}_{e}}\frac{1}{\sqrt{|\Set{N}_{e}||\Set{N}_{e'}|}}\Mat{h}^{(l-1)}_{e'},
\end{align}
where $\Set{N}_{e}=\{e'|(e,e')\in\Set{G}~\text{or}~(e,e')\in\Set{O}^{+}\}$ is the set of nodes which are connected with $e$.
After stacking such $L$ layers, we obtain the final representation for each node $\Mat{h}_{e}=\Mat{h}^{(L)}_{e}$.
Compared with the initial ID embeddings, the GNN model injects graph structure into representation learning, so as to facilitate further exploration.


\subsubsection{\textbf{Neighbor Attention Module.}}\label{sec:neighbor-attention-module}
At state $s_{t}=(u,e_{t})$, having established representations for node $e_{t}$ and its neighbors $\Set{N}_{e_{t}}$, we need to effectively search relevant actions towards potential items.
In particular, we decompose an exploration operation $a_{t}=(e_{t}\rightarrow e'_{t}\rightarrow e_{t+1})$ into two steps:
1) choosing an outgoing edge from $e_t$ to the internal node $e'_t$, \ie $(e_{t},e'_{t})$,
and 2) determining the third node $e_{t+1}$ conditioned on the historical steps.
Such process is as:
\begin{align}\label{equ:policy-probability}
    \Space{P}(a_{t}|s_{t})=\Space{P}((e_{t},e'_{t})|s_{t})\cdot \Space{P}((e'_{t},e_{t+1})|s_{t},(e_{t},e'_{t})),
\end{align}
where $\Space{P}(a_{t}|s_{t})$ represents the confidence or probability of $e_{t+1}$ being negative; $\Space{P}(e_{t},e'_{t})$ and $\Space{P}(e'_{t},e_{t+1})$ separately model the confidence of each exploration step.
Here we implement these two exploration steps via two attention models.

\vspace{5pt}
\noindent\textbf{Attentive KG Neighbors.}
For the current item node $e_{t}$, we need to estimate how likely its related KG entities $\Set{N}_{e_{t}}$ are exposed to user $u$.
Intuitively, a user pays different attentions on various KG entities.
For example, for a movie, a user might care more about the \emph{Director} entity than the \emph{Writer} entity.
This indicates that other movies having the same \emph{Director} entity are more likely to be exposed to the user.
Furthermore, for different movies, her points of interest could change and be dynamic.
For instance, a user watched two movies, because she is attracted by the \emph{Director} and \emph{Star} entities, respectively.
Therefore, we devise an attention model to adaptively specify importance of neighbors, which are sensitive to the current state, \ie user $u$ and current item $e_{t}$.



Formally, for each outgoing edge from $e_{t}$ to neighbor $e'_{t}\in\Set{N}_{e_{t}}$, we generate its representation as $\Mat{h}_{e_{t}}\odot\Mat{h}_{e'_{t}}$, and then include user representation $\Mat{h}_{u}$ to formulate its importance as:
\begin{align}\label{equ:explore-first}
    p(e_{t},e'_{t})=\Trans{\Mat{h}_{u}}\rho(\Mat{h}_{{e}_{t}}\odot\Mat{h}_{e'_{t}}),
\end{align}
where $\odot$ is the element-wise product.
Such importance score is dependent on the affinity between user $u$, item $e_{t}$, and KG entity $e'_{t}$.
Thereafter, we employ a softmax function to normalize the scores across all neighbors as:
\begin{align}\label{equ:explore-first-prob}
    \Space{P}((e_{t},e'_{t})|s_{t})=\frac{\exp{(p(e_{t},e'_{t}))}}{\sum_{e''_{t}\in\Set{N}_{e_{t}}}\exp{(p(e_{t},e''_{t}))}}.
\end{align}
Following such distribution, the KG entity $e'_{t}$ can be selected from the set of candidates $\Set{N}_{e_{t}}$ with the corresponding probability.

\vspace{5pt}
\noindent\textbf{Attention Item Neighbors.}
Having selected KG entity $e'_{t}$, we employ another attention model to decide yield which item from its neighbors $\Set{N}_{e'_{t}}$ as the proposal.
Typically, different unobserved item neighbors have varying confidence of being negative, conditioned on KG entity $e'_{t}$ and user $u$.
Here we model the confidence of each edge from $e'_{t}$ to $e_{t+1}\in\Set{N}_{e'_{t}}$ as:
\begin{align}
    p(e'_{t},e_{t+1})=\Trans{\Mat{h}_{u}}\rho(\Mat{h}_{{e}'_{t}}\odot\Mat{h}_{e_{t+1}}).
\end{align}
Then a softmax function is followed to generate the selection probability of item $e_{t+1}$ as:
\begin{align}\label{equ:explore-second-prob}
    \Space{P}((e'_{t},e_{t+1})|s_{t},(e_{t},e'_{t}))=\frac{\exp{(p(e'_{t},e_{t+1}))}}{\sum_{e''_{t+1}\in\Set{N}_{e'_{t}}}\exp{(p(e'_{t},e''_{t+1}))}}.
\end{align}
As a result, we can generate the negative probability of each exploration operation for a policy $\pi$ (\cf Equation~\eqref{equ:policy}).


\subsubsection{\textbf{Neighbor Pruning Module.}}\label{sec:neighbor-pruning-module}
While such exploration over KG has narrowed the search space from the whole item sets to multi-hop neighbors of positive items, the neighbor scale of some nodes (\eg popular items or general KG concept like the genre of \emph{Drama}) easily reach thousands or even larger.
It is very computational expensive to calculate an output distribution over neighbors of such a node and smooth the probability (via Equations~\eqref{equ:explore-first-prob} and~\eqref{equ:explore-second-prob}), further hindering the exploration performance. Thus, we get inspiration from DNS~\cite{DNS} and propose a pruning strategy that can effectively keep promising neighbors.

More specifically, for each training epoch, we first construct a subset of neighbors, $\hat{\Set{N}}_{e}\subset\Set{N}_{e}$, which consists of $n_{1}$ randomly sampled neighbors (via either sub-sampling or oversampling), so as to reduce the search space.
Moreover, to guarantee the diversity of samples, we additionally introduce some nodes randomly sampled from the whole space.
Thereafter, we design a scoring function to select potential neighbors from $\hat{\Set{N}}_{e}$ to build the set $\tilde{\Set{N}}_{e}\subset\hat{\Set{N}}_{e}$ involving $n_2$ candidate neighbors, heuristically filtering useless nodes out.
Formally, the scoring function is formalized as $g(u,e')=\Trans{\Mat{h}}_{e}\Mat{h}_{e'}$, which is modeled as the representation similarity between $e$ and $e'\in\hat{\Set{N}}_{e}$.
Having obtained the confidence scores, we generate the ranking list over $\hat{\Set{N}}_{e}$ and select the top $n_2$ neighbors to construct $\tilde{\Set{N}}_{e}$.
With such neighbor pruning strategy, we can use $\tilde{\Set{N}}_{e_{t}}$ and $\tilde{\Set{N}}_{e'_{t}}$ to replace the original set $\Set{N}_{e_{t}}$ and $\Set{N}_{e'_{t}}$ in Section~\ref{sec:neighbor-attention-module}, respectively.
As a result, we can effectively solve the high time complexity of neighbor attention module (evidence from Section~\ref{sec:complexity}).

\subsection{Model Optimization}
Finally, we adopt the iteration optimization to train the recommender (\ie MF) and the sampler (\ie KGPolicy), where the recommender and sampler parameters are $\Theta_{R}=\{\Mat{r}_{u},\forall u\in\Set{U},\Mat{r}_{i},\forall i\in\Set{I}\}$ and $\Theta_{S}=\{\Mat{h}^{(0)}_{e},\forall e\in\Set{E},\Mat{W}^{(l)}, \forall l\in\{1,\cdots,L\}\}$, respectively.

\subsubsection{\textbf{Recommender Optimization}}
We first freeze $\Theta_{S}$, sample one negative item $j=e_{T}$ to pair one positive interaction and feed them into the recommender (\ie Equation~\eqref{equ:def-recommender-loss}), and update $\Theta_{R}$ via stochastic gradient descent (SGD)~\cite{BPRMF}.

\subsubsection{\textbf{Sampler Optimization}}
We then freeze $\Theta_{R}$ and update $\Theta_{S}$.
However, SGD cannot be directly applied to do the optimization, since the sampler involves discrete sampling steps which block the gradients when performing differentiation.
A common solution is the policy gradient based RL (REINFORCE)~\cite{PolicyGradient} method, such that the gradients of $\Lapl_{S}$ \wrt $\Theta_{S}$ are calculated as:
\begin{align}
    \nabla_{\Theta_{S}}\Lapl_{S}&=\nabla_{\Theta_{S}}\sum_{(u,i)\in\Set{O}^{+}}\Space{E}_{\pi}[\sum_{t=1}^{T}\lambda^{t-1}\Set{R}(e_{t})]\\
    &\simeq\sum_{(u,i)\in\Set{O}^{+}}\frac{1}{T}\sum_{t=1}^{T}[\gamma^{t-1}R(e_{t})\nabla_{\Theta_{S}}\log{\Space{P}(a_{t}|s_{t})}].\nonumber
\end{align}
Moreover, following prior studies~\cite{NMRN}, we also subtract a baseline $b$ from the policy gradient to reduce the variance.
Wherein, the baseline $b$ is set as the average reward of the recently generated negative interactions.
As such, in each iteration, we alternatively optimize the objective functions of the recommender and sampler.





\subsubsection{\textbf{False Negative Issue}}\label{sec:false-negative}
It is hard for stochastic sampling approaches to avoid the false negative issue --- some items sampled as negative by the sampler during training are truly positive during inference in further test dataset.
Compared to adaptive samplers, KGPolicy benefiting from item knowledge could empirically alleviate the issue to some extent (evidences in Section~\ref{sec:performance-ns}).

\subsubsection{\textbf{Time Complexity Analysis}}\label{sec:complexity}
For the module of graph learning (\cf Section~\ref{sec:graph-learning-module}), establishing representations for nodes has computational complexity $O(\sum_{l=1}^{L}(|\Set{G}|+|\Set{O}^{+}|)d_{l}d_{l-1})$.
For the module of neighbor attention module (\cf Section~\ref{sec:neighbor-attention-module}), the time complexity which mainly comes from the calculation of attention scores is $O(2T|\Set{O}^{+}|\tilde{\Set{N}_{e}|}d^{2})$.
As a result, the time cost of the whole training epoch is $O(\sum_{l=1}^{L}(|\Set{G}|+|\Set{O}^{+}|)d_{l}d_{l-1}+2T|\Set{O}^{+}|n_{2}d^{2})$.
Empirically, RNS, DNS, IRGAN, AdvIR, NMRN, and KGPolicy cost around 22s, 116s, 155s, 175s, 172s, and 232s per training epoch on the largest Yelp2018 dataset, respectively.
Hence, KGPolicy has comparable complexity to adaptive samplers, especially the adversarial ones (IRGAN, AdvIR, and NMRN).






%% file: 4_experiment_correct.tex
\section{Experiment}

We evaluate our proposed KGPolicy on three public datasets, aiming to answer the following research questions:
\begin{itemize}[leftmargin=*]
    \item\textbf{RQ1}: How does KGPolicy perform, compared with existing methods \wrt two dimensions --- the effectiveness of negative sampling and the usage of knowledge entires?

    \item\textbf{RQ2}: How do different components (\eg number of exploration operations, reward functions) affect KGPolicy?
    
    \item\textbf{RQ3}: Can KGPolicy provide in-depth analyses of negative samples?
\end{itemize}

\subsection{Dataset Description}
We use three publicly available datasets: Amazon-book, Last-FM, and Yelp2018, released by KGAT~\cite{KGAT}.
Each dataset is composed of two components, the user-item interactions and KG derived from Freebase (Amazon-book and Last-FM) or local business information network (Yelp2018), as summarized in Table~\ref{tab:dataset}.
Specifically, each knowledge-aware fact is represented as a conceptual edge (item: $i_1$, author: $p_1$).
Note that, we omit the explicit modeling of semantic relations among KG entities and merge the entity type into the entity instances; we leave it for future work.

We use the same training and test sets as that of KGAT~\cite{KGAT}.
That is, for each user, the interaction history is split into the training and test parts with the ratio of $80\%:20\%$.
In the training set, we view each observed user-item interaction as a positive instance, while using the sampler to sample a negative item to pair the same user, and build the recommender.





\subsection{Baselines}
We compare our proposed method, KGPolicy, with two groups of the state-of-the-art baselines, as follows:
\subsubsection{\textbf{Negative Sampling Methods.}}\label{sec:baselines-ns}
To verify the effectiveness of negative sampling, we select the state-of-the-art sampling methods as baselines, covering the static (RNS and PNS), adaptive (DNS, IRGAN, AdvIR, and NMRN), and KG-based (RWS) samplers:
\begin{itemize}[leftmargin=*]
    \item\textbf{RNS}~\cite{BPRMF}: Such \textbf{r}andom \textbf{n}egative \textbf{s}ampling (RNS) is one prevalent technique to sample negative items with uniform probability. For a fair comparison, we use MF as the recommender.
    \item\textbf{PNS}~\cite{PNS,DBLP:conf/kdd/ChenSSH17}: The MF recommender is equipped with the \textbf{p}opularity-biased \textbf{n}egative \textbf{s}ampling (PNS).
    
    \item\textbf{DNS}~\cite{DNS}: This is a state-of-the-art sampling strategy, dynamic negative sampling (DNS), which adaptively picks the negative item scored highest by the current MF recommender among some random missing samples. DNS is known as one of the most effective sampler for BPR loss, and empirically outperforms~\cite{AdaptiveSampler}.
    
    \item\textbf{IRGAN}~\cite{IRGAN}: Such model is an adversarial sampler, which conducts a minimax game between the recommender and the sampler towards selecting better negative items.
    
    \item\textbf{AdvIR}~\cite{AdvIR}: This is a state-of-the-art sampler, which exploits both adversarial sampling and training (\ie adding perturbation) to generate negatives. In particular, the recommender and the sampler use the identical MF models.

    \item\textbf{NMRN}~\cite{NMRN}: This sampler uses the translation-based CF models as the interaction and reward functions in the recommender and sampler, to adversarially sample negatives.
    
    \item\textbf{RWS}~\cite{ACRec}: This \textbf{r}andom \textbf{w}alk \textbf{s}ampling (RWS) depends on the topology of KG merely to select negative items to assist MF.
\end{itemize}

\begin{table}[t]
    \caption{Statistics of the datasets.}
    \vspace{-10px}
    \label{tab:dataset}
    \resizebox{0.42\textwidth}{!}{
    \begin{tabular}{c l|r|r|r}
    \hline
     &  & \multicolumn{1}{l|}{Amazon-book} & \multicolumn{1}{l|}{Last-FM} & \multicolumn{1}{l}{Yelp2018} \\ \hline\hline
    \multirow{3}{*}{\begin{tabular}[c]{@{}c@{}}User-Item\\ Interaction\end{tabular}} & \#Users & $70,679$ & $23,566$ & $45,919$ \\
     & \#Items & $24,915$ & $48,123$ & $45,538$ \\
     & \#Interactions & $847,733$ & $3,034,796$ & $1,185,068$ \\ \hline\hline
    \multirow{3}{*}{\begin{tabular}[c]{@{}c@{}}Knowledge\\ Graph\end{tabular}} & \#Entity Types & $39$ & $9$ & $42$ \\ 
     & \#KG Entities & $88,572$ & $58,266$ & $90,961$ \\
     & \#Edges & $2,557,746$ & $464,567$ & $1,853,704$ \\ \hline
    \end{tabular}}
    \vspace{-15px}
\end{table}

\subsubsection{\textbf{KG-enhanced Recommender Methods.}}\label{sec:baselines-kg}
As KGPolicy presents a new way that uses KG in the sampler, we select the state-of-the-art KG-based recommenders as the competitors, ranging from supervised learning-based (NFM), regularization-based (CKE), path-based (RippleNet) to GNN-based (KGAT):
\begin{itemize}[leftmargin=*]
    \item\textbf{NFM}~\cite{NFM}: This recommender factorizes historical behaviors and item knowledge as the representations of a user-item interaction pair and feeds into a neural network to conduct predictions.
    
    \item\textbf{CKE}~\cite{CKE}: Such recommender uses KG embeddings to enhance item representations and further assist MF.
    
    \item\textbf{RippleNet}~\cite{RippleNet}: Such model leverages multi-hop paths rooted at each user in KG to enrich their representations, and employs MF on representations to do the predictions.
    
    \item\textbf{KGAT}~\cite{KGAT}: This is a state-of-the-art KG-based recommender, which employs GNN on KG to generate representations of users and items and use inner product to do predictions.
\end{itemize}

\subsection{Experimental Settings}
\subsubsection{\textbf{Evaluation Metrics}}
We use two widely-used metrics of top-$K$ recommendation and preference ranking tasks~\cite{NCF,KGAT}: recall@$K$ and ndcg@$K$.
By default, we set $K$ to be $20$.
For each user in the test sets, we view the items she has adopted as positive items, and evaluate how well the recommenders rank the positive items over the whole item space.
We report the average metrics of all users in each test set.

\subsubsection{\textbf{Parameter Settings}}
We implement our KGPolicy model in Pytorch and release our code and datasets at \url{https://github.com/xiangwang1223/kgpolicy}.
As the datasets, data splits, evaluation metrics, and KG-based baselines (\cf Section~\ref{sec:baselines-kg}) are exactly the same as that used in KGAT~\cite{KGAT}, we hence directly copy their performance from the original paper~\cite{KGAT}.
As for KGPolicy and the negative sampling baselines (\cf Section~\ref{sec:baselines-ns}), we fix the embedding size for all recommenders and samplers as $64$, and set the optimizer as Adm~\cite{Adam}.
We use Xavier~\cite{Xarvier} to initialize sampler parameters; meanwhile, as suggested in~\cite{IRGAN,AdvIR}, a trained MF with RNS is used to initialize the recommenders paired with IRGAN, AdvIR, and KGPolicy, so as to stabilize and speed up the model training.
For hyperparameters, we conduct a grid search to find the optimal settings for each model:
the learning rates for the recommender and the sampler are searched in $\{0.0001, 0.0005, 0.001,0.005\}$, and the coefficients of $L_{2}$ regularization in the recommender is tuned in $\{10^{-6}, 10^{-5}, 10^{-4}, 10^{-3}\}$.

Other hyperparameters of KGPolicy are set as follows:
we apply two graph convolutional layers to perform graph representation learning, \ie $L=2$ in Equation~\eqref{equ:gnn-rep};
we search the number of exploration operations, $T$, in $\{1,2,3,4\}$ and report its effect in Section~\ref{sec:impact-k};
moreover, the size of pruned neighbors $n_{2}$ is searched in $\{4,8,16,32,64\}$ where $n_1$ is fixed as $64$.
Without specification, we set $T$ as $2$ by default.

\subsection{Performance Comparison (RQ1)}
We first report the empirical results \wrt negative sampling, and the analyze the comparison \wrt the usage of KG.

\begin{table}[t]
    \caption{Comparison with Negative Samplers.}
    \vspace{-10px}
    \label{tab:overall-performance-ns}
    \resizebox{0.47\textwidth}{!}{
    \begin{tabular}{l|cc|cc|cc}
    \hline
    \multirow{2}{*}{} & \multicolumn{2}{c|}{Yelp2018} & \multicolumn{2}{c|}{Last-FM} & \multicolumn{2}{c}{Amazon-Book} \\
     & recall & ndcg & recall & ndcg & recall & ndcg \\ \hline\hline
    RNS & 0.0465 & 0.0575 & 0.0661 & 0.1063 & 0.1153 & 0.0754 \\
    PNS & 0.0166 & 0.0220 & 0.0668 & 0.0984 & 0.1127 & 0.0730 \\\hline
    DNS & $\Mat{0.0687}$ & $\Mat{0.0839}$ & $\Mat{0.0877}$ & $\Mat{0.1381}$ & $\Mat{0.1518}$ & $\Mat{0.1033}$ \\ 
    IRGAN & 0.0628 & 0.0767 & 0.0642 & 0.1070 & 0.1253 & 0.0833 \\
    AdvIR & 0.0590 & 0.0744 & 0.0810 & 0.1304 & 0.1427 & 0.0967 \\
    NMRN & 0.0565 & 0.0691 & 0.0719 & 0.1151 & 0.1305 & 0.0875 \\ \hline
    RWS & 0.0488 & 0.0611 & 0.0667 & 0.1076 & 0.1185 & 0.0760 \\ \hline
    \textbf{KGPolicy} & $\Mat{0.0747}^*$ & $\Mat{0.0921}^{*}$ & $\Mat{0.0932}^{*}$ & $\Mat{0.1472}^{*}$ & $\Mat{0.1572}^{*}$ & $\Mat{0.1089}^{*}$ \\ \hline\hline
    \%Improv. & 8.73\% & 9.77\% & 6.27\% & 6.59\% & 3.56\% & 5.42\% \\ \hline
    \end{tabular}}
    \vspace{-15px}
\end{table}

\subsubsection{\textbf{Empirical Results \wrt Negative Sampling.}}\label{sec:performance-ns}
Through analyzing the overall results summarized in Table~\ref{tab:overall-performance-ns}, we have the following observations:
\begin{itemize}[leftmargin=*]
    \item Our proposed KGPolicy consistently outperform all baselines across three datasets in all measures. In particular, KGPolicy achieves remarkable improvements over the strongest baselines \wrt ndcg@$20$ by $9.77\%$, $6.59\%$, and $5.42\%$ in Yelp2018, Last-FM, and Amazon-Book, respectively. We attribute such improvements to the following aspects --- 1) by exploiting the rich facts in KG, KGPolicy is more likely to discover high-quality negative items, than the static and adaptive samplers that are guided by limited information; 2) such discovered negatives are close to the target positive interactions, so as to offer meaningful gradients for recommender learning (evidence from the better capacity and representation ability of MF); and 3) by taking advantage of the KG structure and the neighbor pruning strategy, KGPolicy is able to effectively narrow the large-scale search space down to potential items.
    
    \item By jointly analyzing the results across the three datasets, we find that the improvement of KGPolicy on Yelp2018 is the most significant, while that in Amazon-book is the least. This may be caused by the quality of knowledge, since KG in Yelp2018 is constructed by using the local business information and hence is more accurate and targeted as compared to the others.

    \item Static samplers (\ie RNS and PNS) make MF perform poor on three datasets.
    In addition, PNS achieves worse performance than that of RNS in Yelp2018. Such findings are consistent to prior studies~\cite{AdaptiveSampler}, suggesting that the samples from uniform or popularity-biased distribution, are easily discriminated by the recommender, and contribute varnishy (close-to-zero) gradients to update the recommender parameters. This also emphasizes the great need in informative negative instances.
    
    \item Compared with that of RNS and PNS, the results of adaptive samplers (\ie DNS, IRGAN, AdvIR, and NMRN) verify the importance of adaptively selecting negatives. They make the sampling distributions dependent on recommender status and conditioned on the target user.
    
    \item DNS works well on the three datasets. This finding is consistent with previous works~\cite{RNS}. One possible reason is that DNS could effectively reduce the search space via the ranking-aware reject sampling mechanism~\cite{DNS}, suggesting the positive effect of suitable pruning strategy.
    
    \item AdvIR and NMRN substantially outperform IRGAN in most cases. It is reasonable since the key-value attention network and the adversarial perturbations are separately involved in NMRN and AdvIR, which endow the recommender better representation and generalization abilities, respectively. This suggests that, the capacity of sampler have impact on the recommender.
    
    \item While leveraging the identical data to KGPolicy, RWS only achieves comparable performance to the static samplers. It makes sense since the paths generated by random walk usually are biased by the popularity of nodes. Moreover, such sampling also remains unchanged with the recommender status. It again justifies that KGPolicy better leverages KG.
    
\end{itemize}

\begin{table}[t]
    \caption{Comparison with KG-based Recommenders.}
    \vspace{-10px}
    \label{tab:overall-performance-kg}
    \resizebox{0.47\textwidth}{!}{
    \begin{tabular}{l|cc|cc|cc}
    \hline
    \multirow{2}{*}{} & \multicolumn{2}{c|}{Yelp2018} & \multicolumn{2}{c|}{Last-FM} & \multicolumn{2}{c}{Amazon-Book} \\
     & recall & ndcg & recall & ndcg & recall & ndcg \\ \hline\hline
    NFM & 0.0660 & 0.0810 & 0.0829 & 0.1213 & 0.1366 & 0.0914 \\ 
    CKE & 0.0657 & 0.0805 & 0.0736 & 0.1184 & 0.1344 & 0.0885 \\
    RippleNet & 0.0664 & 0.0822 & 0.0791 & 0.1238 & 0.1336 & 0.0910 \\
    KGAT & $\Mat{0.0712}$ & $\Mat{0.0867}$ & $\Mat{0.0870}$ & $\Mat{0.1325}$ & $\Mat{0.1489}$ & $\Mat{0.1006}$ \\ \hline
    \textbf{KGPolicy} & $\Mat{0.0747}^*$ & $\Mat{0.0921}^{*}$ & $\Mat{0.0932}^{*}$ & $\Mat{0.1472}^{*}$ & $\Mat{0.1572}^{*}$ & $\Mat{0.1089}^{*}$ \\ \hline\hline
    \%Improv. & 4.92\% & 6.22\% & 7.12\% & 11.09\% & 5.60\% & 8.25\% \\ \hline
    \end{tabular}}
    \vspace{-15px}
\end{table}

\subsubsection{\textbf{Empirical Results \wrt KG usage.}}
Table~\ref{tab:overall-performance-kg} shows the performance comparison \wrt the usage of KG.
We have the following findings:
\begin{itemize}[leftmargin=*]
    \item Clearly, we can observe the significant improvements brought from KGPolicy on all three datasets \wrt all evaluation metrics. For example, KGPolicy outperforms the strongest baseline, KAGT, \wrt ndcg@$20$ by $6.22\%$, $11.09\%$, and $8.25\%$ on Yelp2018, Last-FM, and Amazon-book, respectively. This again validates the rationality of using KG in the sampling method, and verifies our hypothesis that KG can provide guiding signals towards high-quality negative items.
    
    \item Knowledge-reinforced negatives endows the recommender better representation ability. Specifically, with th except of NFM, all KG-based recommenders user inner product of user and item representations to do predictions; hence the representation ability directly determines how well the recommender perform. Compared with complex representation learning models (\eg path-based in RippleNet and GNN-based in KGAT), KGPolicy uses simple ID embeddings but achieves the best performance. This suggests that, using proper negative signals helps improve representation ability.
   
    \item Existing methods are mainly focusing on using KG to leverage positive signals --- either enhancing semantic similarity among items (\ie CKE), propagating user preference (\ie RippleNet), or encoding high-order connectivity between users and items (\ie KGAT), while ignoring its potential for distilling negative signals. The superior performance of KGPolicy emphasizes that, working together with behavior differences of users, KG is beneficial to negative sampling.
\end{itemize}

\subsection{Study of KGPolicy (RQ2)}
We also perform ablation studies to get deep insights on KGPolicy.
We start by exploring the influence of different KG components --- user behaviors and item knowledge.
We then investigate how the number of exploration operations affects the performance.
In what follows, we also analyze the influence of reward functions.


\subsubsection{\textbf{Impact of Exploration Number}}\label{sec:impact-k}
As the core of KGPolicy is the exploration operation, we hence investigate how the number of such operations affects the performance.
In particular, we search the operation number, $T$, in the range of $\{1,2,3,4\}$.
Table~\ref{tab:impack-num-exploring} summarizes the experimental results, wherein KGPolicy-3 indicates the sampler trained with three exploration operations, similar notations for others.
There are several interesting observations:
\begin{itemize}[leftmargin=*]
    \item Increasing the number of exploration operations enhances the predictive results. Clearly, KGPolicy-3 outperforms KGPolicy-1 and KGPolicy-2 in most cases. We attribute such consistent improvements to the diversity of negative items: three-hop item neighbors derived from KGPolicy-3 naturally cover more items --- beyond these exposed but less interested, and probably involving some unobserved and disliked negatives --- than the one- and two-hop neighbors from KGPolicy-1 and KGPolicy-2.
    
    \item Continuing one more exploration beyond KGPolicy-3, we observe that KGPolicy-4 makes the performance worse across the board. This might be because conducting too many operations would introduce less-relevant items and result in vanishing gradients to the recommender training.
    
    \item Jointly comparing results across Tables~\ref{tab:overall-performance-ns} and~\ref{tab:impack-num-exploring}, KGPolicy with varying exploration operations is superior to other methods consistently. This again empirically shows the rationality and effectiveness of knowledge-aware negative sampling.
\end{itemize}

\subsubsection{\textbf{Impact of Reward Functions}}\label{sec:impact-reward}
To verify the influence of reward functions, we do ablation study by considering two variants of KGPolicy.
In particular, we employ the prediction reward function (\cf~Equation~\eqref{equ:reward}) to build the variant P-Reward, while using the similarity reward function merely to construct the variant P-Reward.
The results of comparison are shown in Table~\ref{tab:impact-reward}.

Only using S+Reward or P+Reward degrades the performance, indicating that they make the selected items suboptimal as negative. There is no exact winner between S-Reward and P-Reward on all datasets. To be more specific, S-Reward performs better in Yelp2018, while outperforming P-Reward in Last-FM. Such slight differences might be caused by the datasets.
Therefore, it validates the rationality of our design.

\begin{table}[t]
    \caption{Impact of exploration operation numbers.}
    \vspace{-10px}
    \label{tab:impack-num-exploring}
    \resizebox{0.45\textwidth}{!}{
    \begin{tabular}{l|cc|cc|cc}
    \hline
    & \multicolumn{2}{c|}{Yelp2018} & \multicolumn{2}{c|}{Last-FM} & \multicolumn{2}{c}{Amazon-Book} \\ 
    & recall & ndcg & recall & ndcg & recall & ndcg \\ \hline\hline
    KGPolicy-1 & 0.0738 & 0.0878 & 0.0891 & 0.1409 & 0.1520 & 0.1053 \\
    KGPolicy-2 & $\Mat{0.0747}^*$ & $\Mat{0.0921}^{*}$ & $\Mat{0.0932}^{*}$ & $\Mat{0.1472}^{*}$ & $\Mat{0.1572}^{*}$ & $\Mat{0.1089}^{*}$  \\
    KGPolicy-3 & 0.0730 & 0.0879 & 0.0928 & 0.1450 & 0.1551 & 0.1076 \\
    KGPolicy-4 & 0.0729 & 0.0878 & 0.0919 & 0.1437 & 0.1546 & 0.1059 \\ \hline
    \end{tabular}}
    \vspace{-10px}
\end{table}

\begin{table}[t]
    \caption{Impacts of reward functions.}
    \vspace{-10px}
    \label{tab:impact-reward}
    \resizebox{0.45\textwidth}{!}{
    \begin{tabular}{l|cc|cc|cc}
    \hline
    & \multicolumn{2}{c|}{Yelp2018} & \multicolumn{2}{c|}{Last-FM} & \multicolumn{2}{c}{Amazon-Book} \\
    & recall & ndcg & recall & ndcg & recall & ndcg \\ \hline\hline
    S-Reward & 0.0731 & 0.0865 & 0.0899 & 0.1411 & 0.1559 & 0.1071 \\
    P-Reward & 0.0721 & 0.0859 & 0.0918 & 0.1443 & 0.1558 & 0.1068 \\ \hline
    \end{tabular}}
    \vspace{-15px}
\end{table}

\subsection{In-depth Analysis (RQ3)}
In this section, we get deep insights on how the knowledge-aware negative sampling facilitates the recommender learning.
Towards the in-depth analysis, we perform additional experiments on the following aspects --- training process \wrt gradient magnitude, and recommendation performance \wrt sparsity levels.
In what follows, we present a case study to illustrate the quality of negative samples.


\subsubsection{\textbf{Training Process \wrt Gradient Magnitude.}}
\begin{figure}[t]
    \centering
    \subfigure[gradient magnitude in Yelp2018]{
    \label{fig:gradient-yelp2018}\includegraphics[width=0.23\textwidth]{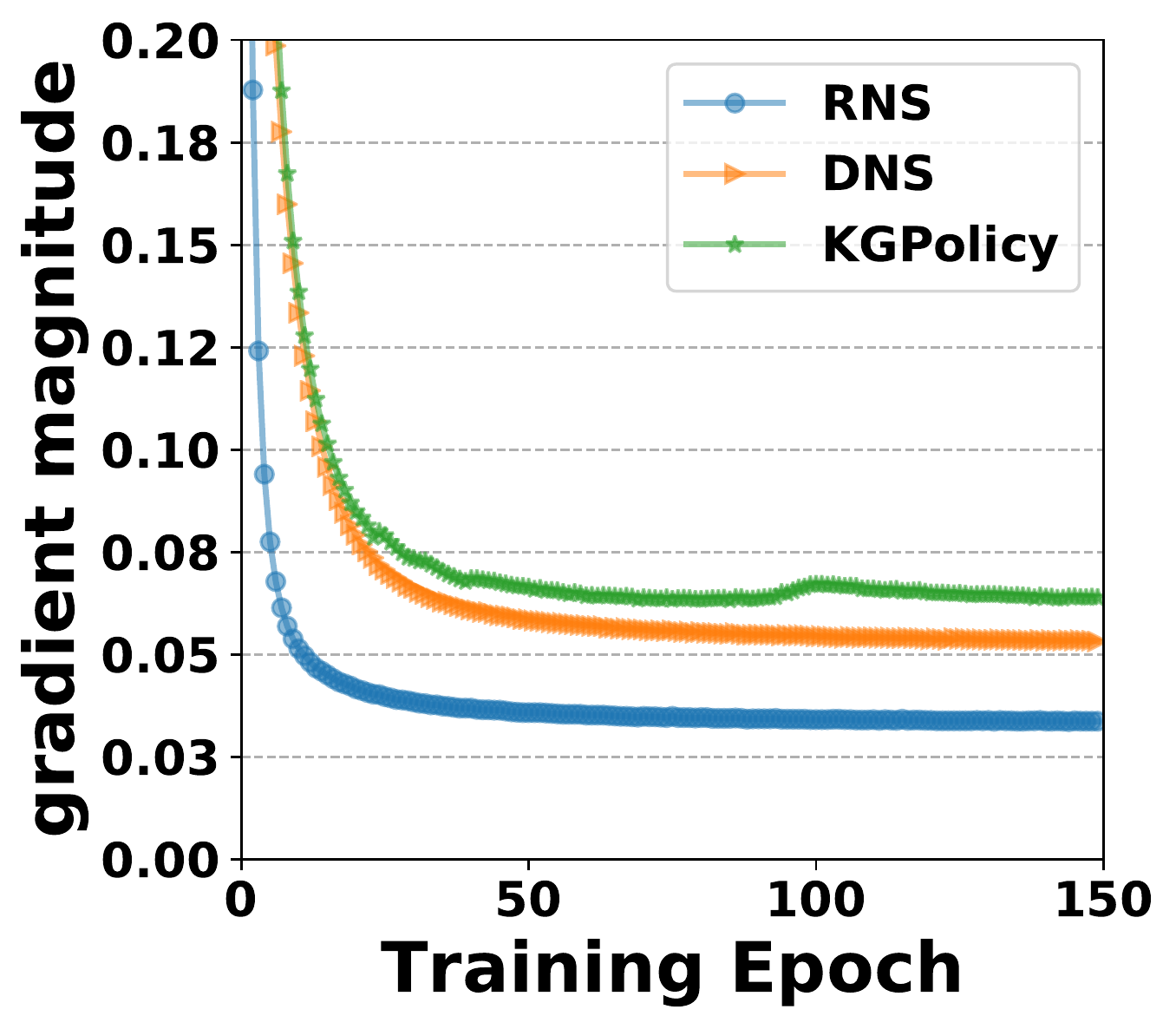}}
    \subfigure[gradient magnitude in Last-FM]{
    \label{fig:gradient-last-fm}\includegraphics[width=0.23\textwidth]{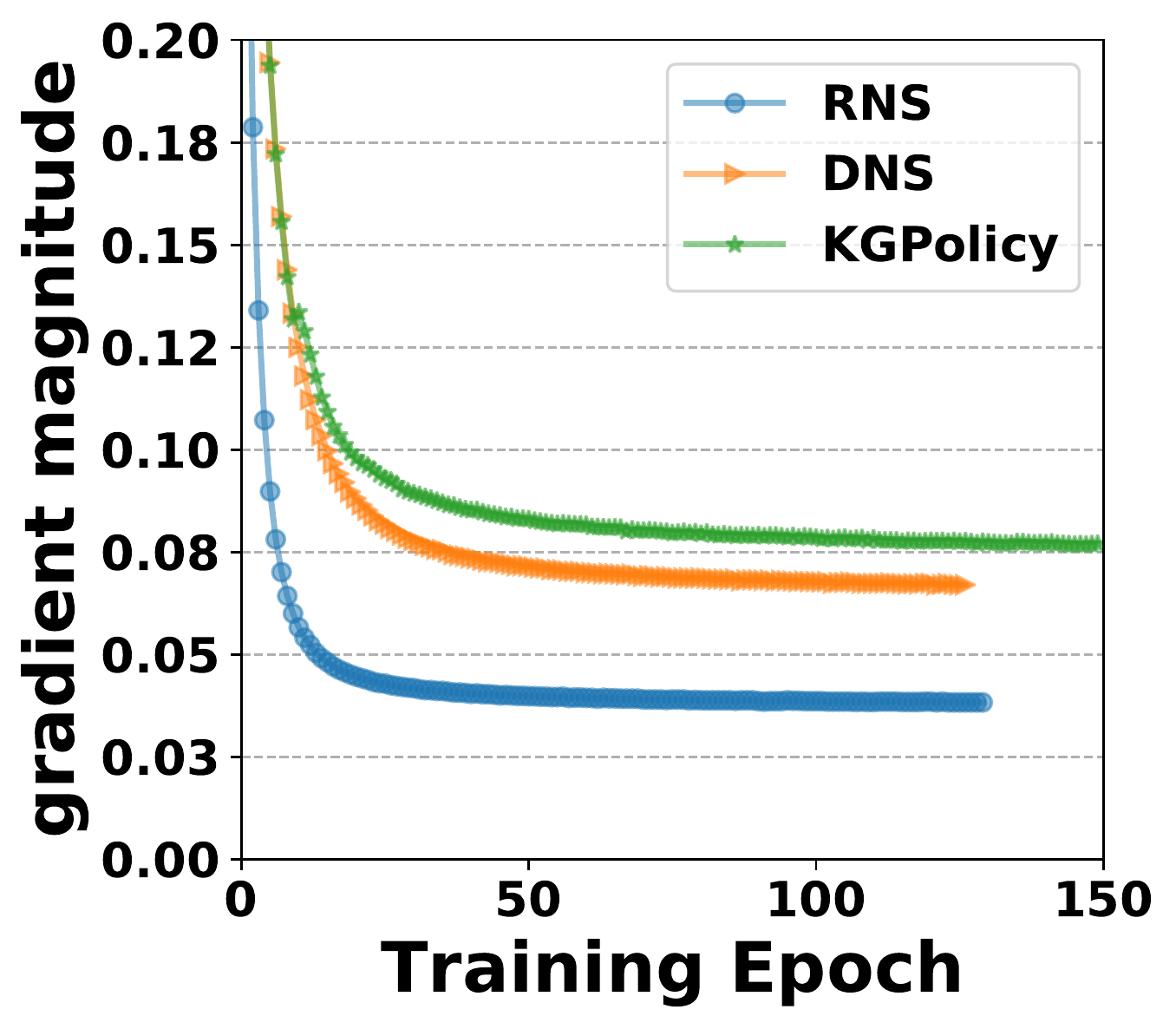}}
    \vspace{-15pt}
    \caption{Training process \wrt average gradient magnitude.}
    \label{fig:gradient}\vspace{-15pt}
\end{figure}

To study how informative the negative items are, we select RNS, DNS, and KGPolicy to represent the static, adaptive, and knowledge-aware sampling approaches, and use the average gradient magnitude in Equation~\eqref{equ:gradient-magnitude} as the evaluation metric.
We record the status of gradient magnitude at each epoch and illustrate the learning curves on Yelp2018 and Last-FM datasets in Figures~\ref{fig:gradient}. Here we omit the similar result in Amazon due to the limited space.

\begin{itemize}[leftmargin=*]
    \item Clearly, knowledge-aware sampling (KGPolicy) achieves larger gradient magnitudes than the static and adaptive strategies throughout all epochs. This qualitative results, together with the performance in Table~\ref{tab:overall-performance-ns}, indicates that the negative items sampled by KGPolicy is more informative than that by RNS and DNS. This again verifies the rationality and effectiveness of incorporating KG into sampling.
    \item The gradient magnitudes of DNS are larger than that of RNS, which is consistent to the observation in~\cite{AdaptiveSampler}. In particular, the uniform sampler easily results in low-quality negative samples, making the gradients varnishing. The suboptimal performance of adaptive sampler emphasizes the importance of knowledge-aware guiding signals.
\end{itemize}

\subsubsection{\textbf{Performance \wrt Sparsity Levels.}}
\begin{figure}[t]
    \centering
    \subfigure[ndcg on Yelp2018]{
    \label{fig:sparsity-yelp2018}\includegraphics[width=0.23\textwidth]{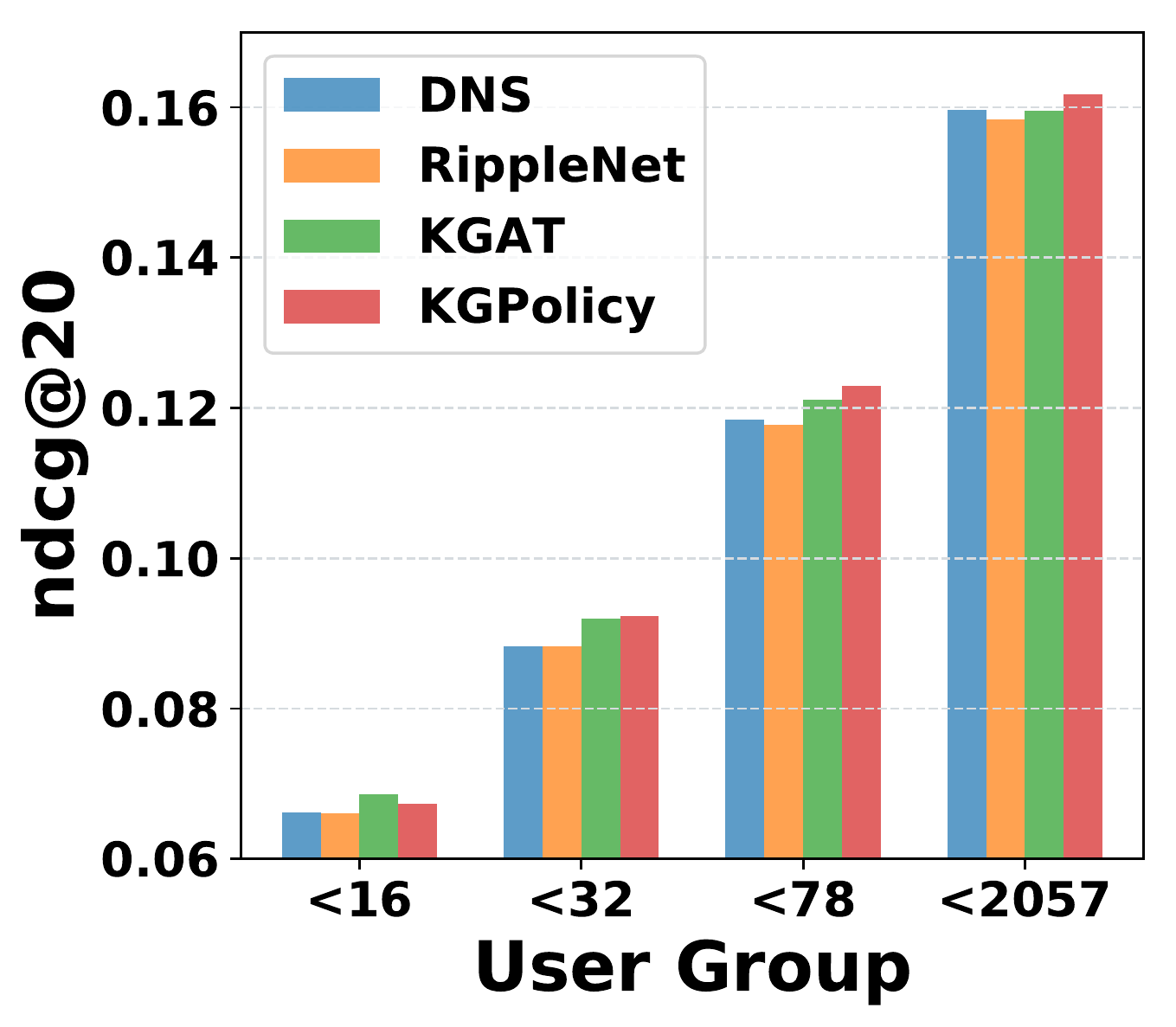}}
    \subfigure[ndcg on Last-FM]{
    \label{fig:sparsity-last-fm}\includegraphics[width=0.23\textwidth]{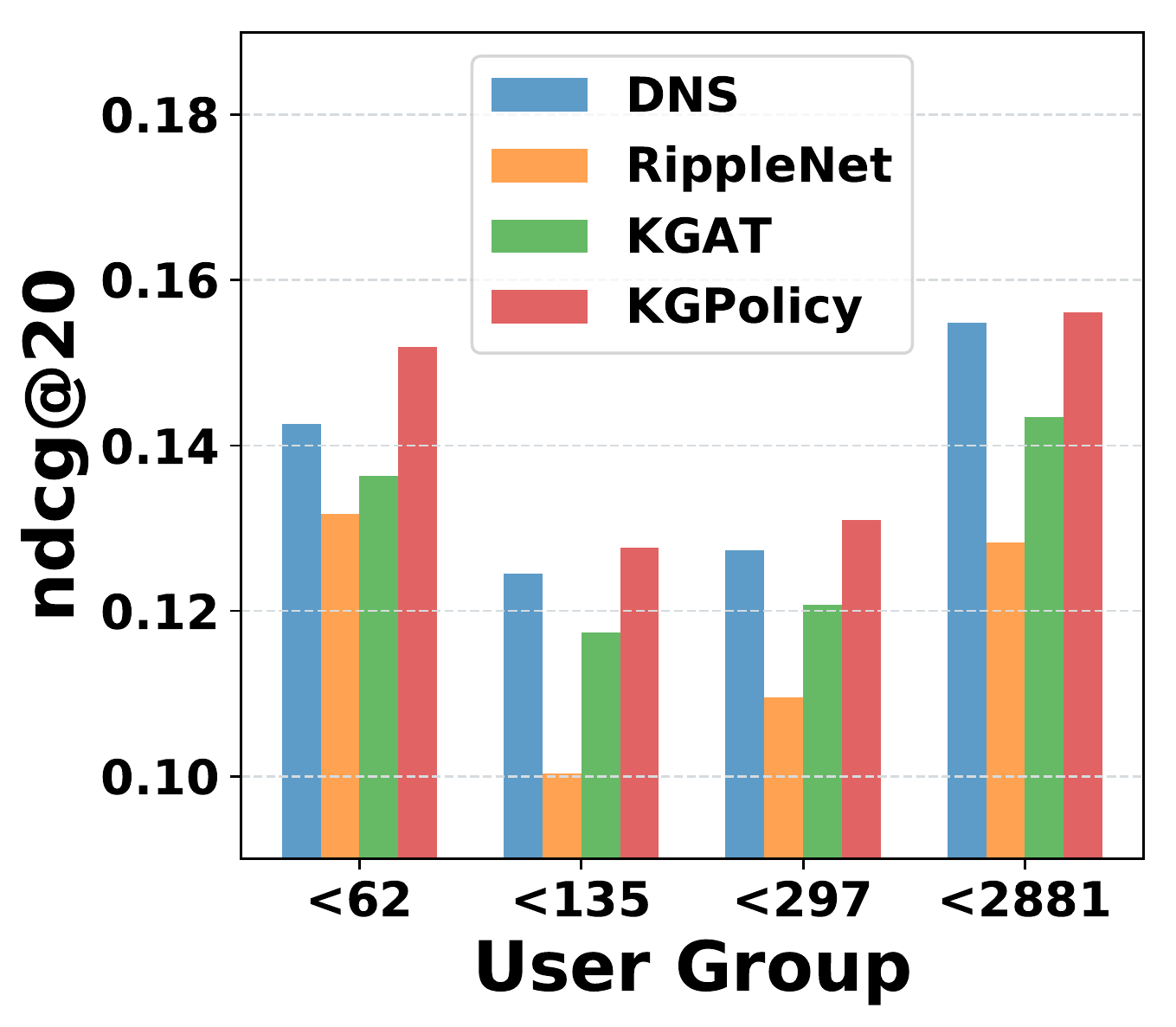}}
    \vspace{-15pt}
    \caption{Performance comparison \wrt sparsity levels.}
    \label{fig:sparsity}\vspace{-10pt}
\end{figure}

Inspired by early works~\cite{CKE,KGAT} which investigates whether KG helps to alleviate the sparsity issue, we would like to track KGPolicy's  contributions on such issue.
Note that, the user groups of different sparsity levels are exactly the same as that reported in KGAT, where users are divided into four groups based on the interaction number per user (\eg separately less than $62$, $135$, $297$, and $2881$ in Last-FM).
We select the strongest baselines \wrt negative sampling (DNS) and KG-based recommendation (RippleNet and KGAT).
Figures~\ref{fig:sparsity-yelp2018} and~\ref{fig:sparsity-last-fm} show their performance \wrt ndcg@$20$ on Yelp2018 and Last-FM, respectively.

\begin{itemize}[leftmargin=*]
    \item Clearly, KGPolicy shows significant improvements over all competing methods in the third and fourth user groups. To be more specific, in Last-FM, the improvements over KGAT are $8.78\%$ and $8.50\%$ for <135 and <297 user groups, respectively. This promising finding again verifies the significance of high-quality negative signals, which can facilitate the simple linear interaction function (\ie MF) to achieve comparable or even better performance than the complex and nonlinear interaction modeling (\eg RippleNet and KGAT).
    
    \item KGPolicy, however, slightly outperforms KGAT in the sparsest group in Yelp2018, where users hold less than $16$ interactions.
    We hence conclude that, knowledge-aware negative sampling is beneficial to the relatively active users, whereas extremely sparse behaviors are not sufficient to guide the sampler learning.
\end{itemize}
In a nutshell, jointly considering informative propagation over KG and knowledge-aware negative sampling might be a promising solution to sparsity issues.
We leave such exploration for future.

\subsubsection{\textbf{Case Study.}}
Exploring paths from the positive interaction to the negative item could be treated as the support evidence why a user holds opposite opinions on two items.
We randomly selected two positive interactions, $(u_{16612},i_{4908})$ and $(u_{23032},i_{21215})$, and illustrate exploring paths derived from KGPolicy-1 in Figure~\ref{fig:case-study}.
We have the following findings:

\begin{figure}[t]
    \centering
	\includegraphics[width=0.42\textwidth]{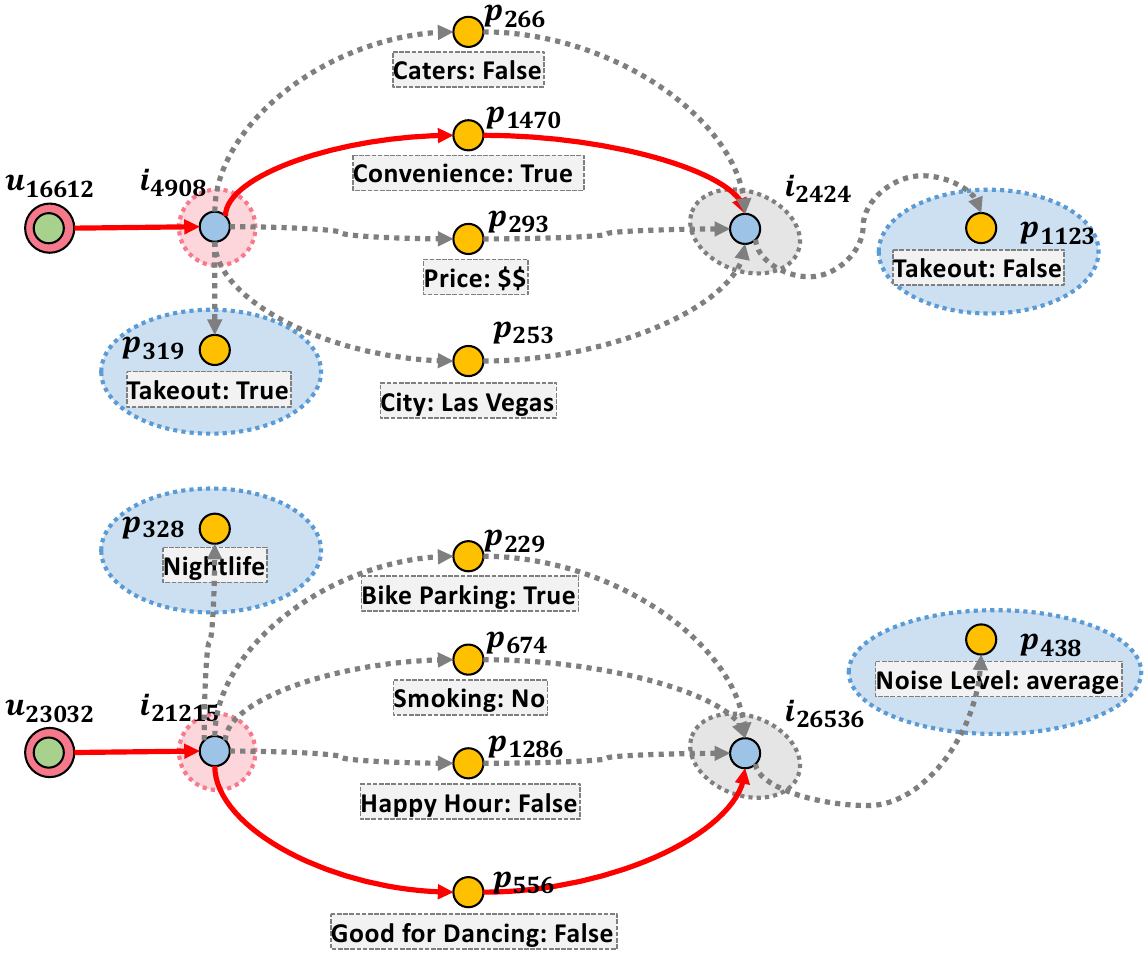}
	\vspace{-10pt}
	\caption{Real cases of exploring paths in Yelp2018.}
	\label{fig:case-study}
	\vspace{-15pt}
\end{figure}

\begin{itemize}[leftmargin=*]
    \item KGPolicy is able to generate exploring paths for each positive interaction, such as $i_{4908}\rightarrow p_{1470}\rightarrow i_{2424}$ for $u_{16612}$ and $i_{21215}\rightarrow p_{1286}\rightarrow i_{26536}$ for $u_{23032}$. Such exploring paths play a crucial role in negative sampling.
    
    \item Taking advantages of the designed neighbor attention module, KGPolicy successfully captures personal interest for each user, specifying varying importance of KG entities. There are many shared neighbors between the positive and negative items. For example, $u_{16612}$ cares more about whether visiting a restaurant is convenient or not, rather than other general attributes; while, $u_{23032}$ prefers the places with the KG entity $p_{556}$, \emph{Good for Dancing: False}. As such, KGPolicy can narrow the searching space down to a set of candidates meeting personal tastes.
    
    \item Furthermore, the exploring paths, together with entity differences between positive and negative items, provides us with some insights of user real tastes. For example, when being aware of the overlapping attributes (\eg $p_{266}$, $p_{1470}$, $p_{293}$, $p_{253}$), $u_{16612}$ visited $i_{4908}$, but ignore $i_{2424}$. Analyzing the attribute differences, we find that $i_{4908}$ allows \emph{TakeOut} (\cf $p_{319}$), which is disallowed in $i_{2424}$. Analogously, exclusive KG entities of $i_{21215}$ are highly likely to make $u_{23032}$ ignores the alternative $i_{26536}$. Therefore, KGPolicy can help interpret why a user consumes the positive items but did not adopt the negatives.
    
    \item These observations also inspires us to involve some true negative signals, towards better explanations on user behaviors. We would like to perform user experience modeling in future work.

\end{itemize}

%% file: 5_related.tex
\section{Related Work}
This work is relevant to two research lines: negative sampling and knowledge-graph based recommendation.

\subsection{Negative Sampling for Recommendation}
Towards solving one-class problem in implicit feedback, earlier recommendation methods~\cite{BPRMF,RNS,PNS,OCCF,DBLP:conf/www/ChengDZK18} use negative sampling to subsample some from unobserved items as negatives, based on the predefined sampling distributions.
For example, random (RNS)~\cite{RNS,BPRMF} and popularity-biased (PNS)~\cite{PNS,DBLP:conf/kdd/ChenSSH17} sampling are based on the uniform and popularity distribution, respectively.
While being prevalent, such static sampling is independent of model status and unchanged for different users, thereby easily low-quality negative items and contributing little to recommender training.
Later on, adaptive sampling, such as DNS~\cite{DNS}, LambdaFM~\cite{LambdaFM}, and Adaptive Oversampling~\cite{AdaptiveSampler}, is proposed.
The basic idea is to devise additional measures to select hard negative items, accounting for model status.
For example, DNS picks the item with the highest prediction scored by the current recommender as a negative sample.
Recent studies~\cite{IRGAN,NMRN,AdvIR} get inspiration from generative adversarial learning~\cite{GAN} to generate adversarial samples.
For example, IRGAN~\cite{IRGAN} and NMRN~\cite{NMRN} introduce a sampler model to play a minimax game with the recommender, so as to select close-to-observed negatives.
While being effective from the perspective of numerical optimization, these hard negative samples are highly likely to be true positive in future testing data.
Hence, more informative guiding signals are required to discover true negative feedback.
Some recent efforts~\cite{DBLP:conf/www/DingF0YLJ18,RNS,DBLP:conf/recsys/LoniPLH16} consider extra behavior data to enhance the negative sampler.
For instance, RNS~\cite{RNS} and multi-channel feedback BPR~\cite{DBLP:conf/recsys/LoniPLH16} exploit viewed but non-clicked and clicked but non-purchased data to help filter negatives.
However, the scale of such behaviors is limited compared with the vast amount of missing data, which is insufficient to effectively distill negative signals.

Distinct from these methods, we hypothesize that knowledge graph (KG) of user behaviors and item knowledge is useful to infer informative and factual negative items from missing data.

\subsection{Knowledge Graph-based Recommendation}
KG-based recommendation has attracted increasing attention.
At its core is the KG-enhanced interaction modeling, which use the structural knowledge to enrich relations among users and items in the recommender.
A research line~\cite{CKE,DKN,KGMemory18,NFM} uses KG embeddings to improve the quality of item representations.
For example, CKE~\cite{CKE} exploits KG embeddings to enhance MF, while DKN~\cite{DKN} treat semantic embeddings from KG as the content information of items.
Another line extracts paths and meta paths~\cite{MCRec,KGRnn18,KPRN} that connect the target user and item via KG entities, and then build the predictive models upon them.
More recently, PGPR~\cite{PGPR} further exploits a RL approach to explore items of interest for a target user.
Another type of methods unify the foregoing ideas to encode first-order~\cite{KTUP,CFKG} and higher-order connectivity~\cite{KGAT,KGCN} of KG into the representations of users and items.
For example, KTUP~\cite{KTUP} and CFKG~\cite{CFKG} jointly train the recommendation task with the KG completion problem, such that first-order connectivity involved in KG triples and user-item pairs can be captured via translation principles.
Taking advantages of information propagation proposed by GNNs~\cite{GCN,GAT}, KGAT~\cite{KGAT} and KGNN-LS~\cite{KGCN} perform embedding propagation by applying multiple GNN layers over KG, so as to directly inject high-order connectivity into the representations.

However, to the best of our knowledge, existing recommenders only leverage KG to design more complex interaction functions and essentially distill better positive signals, but leave the negative signals unexplored.
Our work differs from them in that, we focus on knowledge-aware negative sampling, towards discovering informative and factual negative feedback from missing data.

%% file: 6_conclusion.tex
\section{Conclusion and Future Work}
In this work, we initiated an attempt to incorporate knowledge graph into negative sampling to discover high-quality negative signals from implicit feedback.
We devised a new knowledge-aware sampling framework, KGPolicy, which works as a reinforcement learning agent and effectively learns to navigate towards potential negative items for a positive user-item interaction.
At its core is the proposed exploration operation, which is capable of adaptively selecting next neighbors to visit, accounting for user behavior- and item knowledge-based negative signals.
Extensive experiments on three benchmark datasets demonstrate the rationality and effectiveness of knowledge-aware sampling.
In particular, when equipped with KGPolicy, MF, such a simple and linear model exhibits significant improvements over both state-of-the-art sampler and KG-based recommender models.

Analysis on how knowledge graph facilitates the recommender learning provides us with some insights of the sampling process.
Distilling high-quality negative signals from unobserved data is of crucial importance to make the use of positive-only data effectively.
It greatly helps to establish better representations of users and items with limited data.
Moreover, such guiding signals sort of interpret user intents and devise shaper preference distributions. 
This work hence opens up new research possibilities.
In future work, we would like to involve more auxiliary information, such as social network and contextual information (\eg geo location and timestamp), together with user behaviors, to better mine negative user feedback.
Furthermore, we plan to perform experiments on user experience modeling, with target of establishing ground truth on what a user likes and dislikes, as well as user-generated explanations.
Exploiting such interpretable and explicit negative signals is beneficial to explainable recommendation.

\begin{acks}
    This research is part of NExT++ research, which is supported by the National Research Foundation, Prime Minister's Office, Singapore under its IRC@SG Funding Initiative, and also supported by the National Natural Science Foundation of China (61972372, U19A2079).
\end{acks}